\documentclass[sn-mathphys-ay]{sn-jnl}

\usepackage{graphicx}%
\usepackage{multirow}%
\usepackage{amsmath,amssymb,amsfonts}%
\usepackage{amsthm}%
\usepackage{mathrsfs}%
\usepackage[title]{appendix}%
\usepackage{xcolor}%
\usepackage{textcomp}%
\usepackage{manyfoot}%
\usepackage{booktabs}%
\usepackage{algorithm}%
\usepackage{algorithmicx}%
\usepackage{algpseudocode}%
\usepackage{listings}%

\usepackage{bm}
\usepackage{tikz}
\usetikzlibrary{decorations.pathreplacing}
\usepackage{multicol}

\theoremstyle{thmstyleone}%
\theoremstyle{thmstyletwo}%

\theoremstyle{thmstylethree}%

\begin{document}

\title[Assessing the Impact of Block Size on Block Likelihood Estimation: A Comparative Study]{Assessing the Impact of Block Size on Block Likelihood Estimation: A Comparative Study}


\author{\fnm{Alfredo} \sur{Alegr\'ia}}\email{alfredo.alegria@uc.cl}

\affil{\orgdiv{Department of Statistics}, \orgname{Pontificia Universidad Cat\'olica de Chile}, \orgaddress{\street{Avenida Vicu\~na Mackenna 4860}, \city{Santiago}, \country{Chile}}}


\abstract{This paper focuses on block likelihood estimation for geostatistical data, a method that balances statistical accuracy and computational efficiency. Central to this approach is the choice of block size, which can significantly impact performance. This study contributes by providing a  thorough numerical investigation of the effects of large versus small block configurations.  Findings from both simulation experiments and real-data analyses of sea surface temperature challenge the prevailing assumption that larger block sizes invariably lead to improved statistical performance.}

\keywords{Cauchy covariance, Composite likelihood, Gaussian process,  Godambe information, Large spatial datasets, Matérn covariance}


\pacs[MSC Classification]{62M40, 62F10}

\maketitle

\section{Introduction}

Gaussian processes are fundamental for analyzing geostatistical data across diverse domains, including climatology, environmental sciences, geology, and ecology \citep{chiles2012geostatistics, cressie2015statistics}. Regardless of the  specific context, accurately estimating the covariance function is essential both for characterizing the underlying dependence structure and for obtaining precise predictions at unsampled locations.

In the contemporary era, the availability of large volumes of data provides valuable opportunities to advance statistical methodologies, yet also introduces substantial computational challenges.
 Specifically, when employing maximum likelihood estimation, a gold standard in statistics, it is customary to use Cholesky factorization as an intermediate step for computing the determinant and inverse of the covariance matrix. However, the computational complexity of this procedure scales cubically with the sample size, motivating researchers to develop more efficient methodologies, making it a central focus of ongoing research \citep{heaton2019case,huang2021competition,hong2023third}. 

A variety of techniques utilize matrices with constrained structures to approximate the covariance matrix. These methods include the use of low-rank approximations \citep{banerjee2008gaussian}, hierarchical matrices \citep{litvinenko2020hlibcov}, sums of Kronecker products for gridded observations \citep{cao2021sum}, and covariance tapering. The latter employs compactly supported covariance functions, producing sparse covariance matrices that are amenable to specific numerical algorithms \citep{furrer2006covariance, kaufman2008covariance, shaby2012tapered}.

Efficient computation can also be attained by approximating the inverse covariance matrix via the precision matrix of a Gaussian Markov random field \citep{lindgren2011explicit}. Additional approaches include using Monte Carlo simulations to approximate score equations \citep{10.1214/13-AOAS627}, exploring modified unbiased estimating equations \citep{sun2016statistically}, and applying spectral techniques, which enable computationally tractable approximations of the log-likelihood \citep{whittle1954stationary, fuentes2007approximate}.

The Vecchia approximation  \citep{vecchia1988estimation} employs product of conditional sub-likelihoods with small conditioning sets, serving as a substitute for the full likelihood. This approach considers conditional densities involving neighboring observations, enabling efficient computation while retaining essential statistical information. For further insights and extensions, see \cite{stein2004approximating} and \cite{10.1214/19-STS755}.

Composite likelihood (CL) methods \citep{lindsay1988composite, varin2011overview}  utilize products of likelihoods corresponding to marginal or conditional events.  CL estimation has been extensively studied in spatial models; see, e.g.,  \citet{curriero1999composite}, \citet{caragea2007asymptotic}, \citet{bai2012joint}, \citet{bevilacqua2012estimating}, \citet{eidsvik2014estimation}, \citet{bevilacqua2016composite}, and \citet{ruli2016approximate}. CL methods are appealing due to their balance of analytical simplicity  and computational tractability. They are also highly flexible, as shown by their effectiveness with irregularly spaced data and the fact that they require specifying only low-dimensional distributions.

We focus on a variant of CL known as block likelihood. In this framework, the sample is partitioned into blocks, after which pairwise block log-likelihoods are constructed and combined additively. Some methods employ medium- to large-sized blocks, containing hundreds to thousands of observations \citep{eidsvik2014estimation}. At one extreme, treating the entire sample as a single block recovers the full likelihood; at the other, blocks may consist of a single observation, yielding the pairwise likelihood and representing the minimal block size.

In this work, we contrast the performance of block likelihood methods with respect to the block sizes employed. Within the small-block segment, we consider the pairwise likelihood method and introduce an additional approach based on blocks of size two that are combined through conditioning. Incorporating this variant extends the analysis beyond the pairwise approach alone, allowing for more thorough investigations within the small-block framework. 

Our simulation studies reveal that, in certain scenarios, substantial increases in block size do not consistently yield proportional statistical improvements. These findings are further demonstrated through an application to sea surface temperature data provided by NOAA PSL in Boulder, Colorado, USA.

The article is organized as follows. Section \ref{sec:preliminaries} provides a brief overview of the challenges associated with maximum likelihood estimation for Gaussian spatial processes, along with a general perspective on CL methods. Section \ref{sec:proposal} introduces the block likelihood approach and discusses variants based on small-sized blocks. Simulation studies are presented in Section \ref{sec:simulations}, and a real data application is illustrated in Section \ref{sec:data}. The paper concludes with a discussion in Section \ref{sec:conclusions}.

\section{Preliminaries}
\label{sec:preliminaries}

\subsection{Likelihood Inference for Gaussian Processes}

Let $\left\{Z(\bm{s}): \bm{s}\in\mathbb{R}^d\right\}$ be a real-valued Gaussian process, such that $\text{var}\left( Z(\bm{s}) \right) < \infty$ for all $\bm{s}\in\mathbb{R}^d$. We assume that the process has zero mean, indicating prior detrending of the observed data. Moreover, in numerous real-world scenarios, correlations decline as distances increase, a property referred to as isotropy \citep{chiles2012geostatistics}. To account for this behavior, we assume that, for any pair of sites $\bm{s},\bm{s}'\in\mathbb{R}^d$, the covariance function of the process can be expressed as 
$$\text{cov}\left( Z(\bm{s}),Z(\bm{s}')\right)= \sigma^2 \rho(h;\bm{\phi}),$$
where $h = \|\bm{s}-\bm{s}'\|$ represents the Euclidean distance between the locations. The function $\rho(\cdot;\bm{\phi})$ denotes a correlation function dependent on a vector of parameters $\bm{\phi} \in \mathbb{R}^q$. These parameters allow us to regulate important aspects, such as the shape and rate of decay of the correlation structure. Here, $\sigma^2>0$ is an additional parameter enabling control over the variance of the process.

To enhance the model's versatility, it is common to incorporate a discontinuity in the covariance function at the origin, known as the nugget effect \citep{chiles2012geostatistics}. This extra  element proves valuable in addressing the  occurrence of measurement errors that impact experimental data. The magnitude of this effect is regulated by a non-negative parameter, denoted as $\tau^2$. Consequently, the covariance function takes the form $\sigma^2 \rho(h;\bm{\phi}) + \tau^2 1\{h=0\}$, where $1\{\cdot\}$ represents the indicator function.

We assume that the process has been sampled at $n$ locations $\bm{s}_1, \ldots, \bm{s}_{n}$. We use the notation $Z_i = Z(\bm{s}_i)$ and consider the vector of data $\bm{Z} = \left(Z_1, \ldots, Z_n\right)^\top$, where $\top$ denotes \emph{transpose}. Then, $$\bm{Z}\sim \mathcal{N}_n(\bm{0},\sigma^2 \bm{R}(\bm{\phi}) + \tau^2 \bm{I}_n),$$ where $\bm{I}_n$ is the $n\times n$ identity matrix, and the $(i,j)$-th entry of $\bm{R}(\bm{\phi})$ is given by $\rho\left(\|\bm{s}_i-\bm{s}_j\|; \bm{\phi}\right)$, for each $i,j=1, \ldots, n$.

The estimation of the parameters can be performed by maximizing the log-likelihood function, which, up to a positive constant, is given by
\begin{equation}
\label{loglik}
\ell(\tau^2,\sigma^2,\bm{\phi}) = - \frac{1}{2} \left( \log \big|\sigma^2 \bm{R}(\bm{\phi}) + \tau^2 \bm{I}_n \big| + \bm{Z}^\top \left[\sigma^2 \bm{R}(\bm{\phi}) + \tau^2 \bm{I}_n \right]^{-1} \bm{Z}\right),
\end{equation}
where $|\cdot|$ denotes the determinant. 

Evaluating the determinant and the inverse in (\ref{loglik}) poses a substantial challenge for scientists dealing with large sample sizes, involving $O(n^{2.81})$ steps (see, e.g., \citealp{aho1974design}). Traditional algorithms such as Cholesky decomposition have a complexity of $O(n^3)$. This challenge has prompted scientists to explore new statistical methodologies for conducting inferences.

\subsection{Composite Likelihood}

The composite likelihood (CL) estimation framework (refer, for example, to  \citealp{lindsay1988composite,varin2011overview}) considers the following paradigm: provided a set of marginal or conditional data segments $A_k$, $k=1,\hdots,K$, along with their corresponding (sub) log-likelihood functions $\ell_k(\bm{\theta})$, the CL estimates are obtained by maximizing \textcolor{black}{the composite log-likelihood function}
\begin{equation}
\label{CL}
\text{\textcolor{black}{cl}}(\bm{\theta}) =
\sum_{k=1}^K \omega_k  \,  \ell_k(\bm{\theta}).
\end{equation}
Here, for the sake of simplicity, we denote the vector of parameters as $\bm{\theta} = (\tau^2, \sigma^2, \bm{\phi}^\top)^\top$, while $\omega_1, \hdots, \omega_K$ represent weights designed to enhance the statistical and computational efficiency of the method. This approach is a misspecified likelihood, where the misspecification originates from assuming independence among segments. Methods in this class offer a balance between statistical and computational efficiency, as the segments of data yield sub-likelihoods that are more computationally cost-effective.

Given that (\ref{CL}) is expressed as the sum of  log-likelihoods, it serves as an unbiased estimation equation. Assuming adequate regularity conditions \citep{varin2008composite}, the CL estimator is consistent and asymptotically Gaussian under increasing domain asymptotics (i.e., when the spatial domain expands with the increasing number of observations). In such a context, we must consider the Godambe information matrix, defined as
\begin{equation*}
\label{godambe_matrix}
{G}(\bm{\theta}) = H(\bm{\theta}) J(\bm{\theta})^{-1}H(\bm{\theta}),
\end{equation*}
where  $H(\bm{\theta}) = {E}\left( - \nabla^2 \text{CL}(\bm{\theta}) \right)$ and $J(\bm{\theta}) = {E}\left( \nabla \text{CL}(\bm{\theta}) \nabla \text{CL}(\bm{\theta}) ^\top \right)$, with $\nabla^2$ and $\nabla$ representing the Hessian and gradient, respectively. The asymptotic variance of the CL estimator is given by the inverse of the Godambe information matrix.

\textcolor{black}{Notice that there is another type of asymptotic framework, termed fixed domain (or infill) asymptotics, which considers a fixed region within which the number of sampling locations increases. The behavior of the estimators can differ substantially between these two asymptotic frameworks. For a detailed comparison, the reader is referred to \cite{zhang2005towards}.}

In this work, the primary motivation for exploring CL methods is to achieve computational efficiency in Gaussian process models. In other settings, however, CL methods are motivated by the analytical difficulty of deriving likelihood functions in closed form. This challenge is particularly pronounced for certain classes of non-Gaussian processes, whose finite-dimensional distributions are inherently intricate. For further discussion of CL methods applied to spatial data, especially in contexts involving departures from Gaussianity, such as skewed distributions, binary outcomes, or the analysis of extremes, we refer the reader to \cite{heagerty1998composite}, \cite{sang2014tapered}, \cite{castruccio2016high}, and \cite{alegria2017estimating}.

\section{Block Likelihood Estimation}
\label{sec:proposal}

\subsection{General Formulation}

We restrict our attention to a specific category of CL methods, referred to as block composite likelihood \textcolor{black}{(referred to by the acronym BCL)}. This approach, proposed by \cite{eidsvik2014estimation}, involves dividing the data into $m$  pieces. A partition $\Delta = \{b_1,\hdots,b_m\}$ of the set $\{1,\hdots,n\}$ is considered, where the $i$-th block of data is defined as $\bm{Z}_{b_i} = (Z_j: j\in b_i)$, for $i=1,\hdots,m$. The cardinality of $b_i$ is denoted by $m_i$.

\cite{eidsvik2014estimation} considered
 pairs of blocks $\bm{Z}_{b_i,b_j} = (\bm{Z}_{b_i}^\top,\bm{Z}_{b_j}^\top)^\top$, each with a corresponding log-likelihood 
 \begin{align*}
 \mathcal{\ell}_{b_i,b_j}(\tau^2,\sigma^2,\bm{\phi})  = & -\frac{1}{2} \big(  \log \big|\sigma^2 \bm{R}_{b_i,b_j}(\bm{\phi}) + \tau^2\bm{I}_{m_i+m_j} \big|  
  \\ &+  \bm{Z}_{b_i,b_j}^\top [\sigma^2\bm{R}_{b_i,b_j}(\bm{\phi}) + \tau^2\bm{I}_{m_i+m_j}]^{-1}\bm{Z}_{b_i,b_j} \big), 
 \end{align*}
 where $\bm{R}_{b_i,b_j}(\bm{\phi})$ denotes the sub-matrix of $\bm{R}(\bm{\phi})$ associated to $\bm{Z}_{b_i,b_j}$.

 The objective function of this method is given by the \textcolor{black}{block composite log-likelihood}
\begin{equation*}\label{eq:block}
\text{\textcolor{black}{bcl}}_m(\tau^2,\sigma^2,\bm{\phi}) =  \sum_{i=1}^{m-1} \sum_{j=i+1}^m \omega_{ij}  \mathcal{\ell}_{b_i,b_j}(\tau^2,\sigma^2,\bm{\phi}).
\end{equation*}
The subscript $m$ highlights the method's dependence on the number of blocks. When $m=1$ or $m=2$, the full likelihood is recovered.  Additionally, there is a significant dependence on the chosen configuration for the blocks $\Delta$; however, we omit this to prevent the notation from becoming too intricate.

A common strategy is to use 0/1 weights for $\omega_{ij}$, that is, only pairs of neighboring or nearby blocks are considered. Proximity can be measured, for example, by the Euclidean distance between their centers of mass (see \citealp{eidsvik2014estimation}).

\subsection{Block Likelihood Based on Small-Sized Blocks}
\label{sec:small}

Since this work, in part, aims to highlight the competitiveness of variants based on small-sized blocks, we provide below a detailed description of two alternatives within that segment.

\subsubsection*{Pairwise Likelihood}

Consider $n$ blocks of the form $b_i = \{i\}$, that is, each block consists of a single observation. This construction yields the pairwise CL (referred to by the acronym PCL)  method, for which the objective function simplifies to
\begin{align}
\label{pcl}
\text{\textcolor{black}{pcl}}(\sigma^2,\bm{\phi}) = & -\frac{1}{2}\sum_{i=1}^{n-1} \sum_{j=i+1}^n \omega_{ij} \bigg[  \log\left( \left(\sigma^2+\tau^2\right)^2 - \left(\sigma^2 \rho_{ij}\right)^2\right) \\ \nonumber
 & + \frac{ \left(\sigma^2+\tau^2 \right)\left( Z_i^2 + Z_j^2 \right)  - 2 \sigma^2 \rho_{ij}Z_iZ_j }{(\sigma^2+\tau^2)^2 - (\sigma^2 \rho_{ij})^2} \bigg],
\end{align}
where $\rho_{ij} = \rho(\|\bm{s}_i-\bm{s}_j\|; \bm{\phi})$.  

A notable feature of PCL is that it obviates the need to manipulate matrices, given the explicit expression (\ref{pcl}). Although other CL versions based on conditioning or differences of pairs exist \citep{curriero1999composite,bevilacqua2015comparing}, we focus on (\ref{pcl}), as other versions generally exhibit equal or inferior performance, as documented in \cite{bevilacqua2015comparing}.

As distant pairs of data tend to exhibit weaker correlations, their contribution to the estimation process is relatively limited. Hence, it is common to employ a 0/1 structure for $\omega_{ij}$. More precisely, 
$$\omega_{ij} = 
\begin{cases}
1 & \text{ if } \|\bm{s}_i-\bm{s}_j\| < d_s\\
0 & \text{ otherwise }
    \end{cases},$$
where $d_s > 0$ is a predefined threshold. This not only improves computational performance but also has the potential to enhance statistical efficiency. This concept has been extensively discussed in the literature (see, e.g., \citealp{caamano2024nearest}). 

\textcolor{black}{
As an alternative to the previous weighting scheme, \citet{caamano2024nearest} consider weights of the form 
\begin{equation}
\label{peso_asim}
    \omega_{ij}(k) = \begin{cases}
1 & \text{ if } \bm{s}_i \in N_k(\bm{s}_j) \\
0 & \text{ otherwise},
\end{cases} 
\end{equation}
where $N_k(\bm{s}_j)$ denotes the set of the $k$ nearest neighbors of site $\bm{s}_j$. Depending on the spatial distribution of the locations, these weights may be asymmetric by construction; see \citet[Figure~1]{caamano2024nearest} for an illustration. This approach can incorporate pairs of sites that distance-based weights may ignore, thereby capturing additional information in the objective function.
}

\subsubsection*{Bi-Conditional Likelihood}

 We start by introducing notation to facilitate the formulation of the objective function. Let $\Omega = \{\bm{s}_1, \ldots, \bm{s}_{n}\}$ denote the set of locations. Assume $n$ is even (in the case of odd $n$, one location must be excluded initially). We partition the set $\Omega$ into $\Omega = \bigcup_{i=1}^{n/2} \Omega_i$, where each subset $\Omega_i$ comprises two sites, $\bm{s}_i^a$ and $\bm{s}_i^b$, for all $i=1, \ldots, n/2$. We define $Z_i^a = Z\left(\bm{s}_i^a\right)$, $Z_i^b = Z\left(\bm{s}_i^b\right)$, and $\bm{Z}_i = (Z_i^a, Z_i^b)^\top$. We also define $\rho_{ij}^{ab} = \rho(\|\bm{s}_i^a - \bm{s}_j^b\|; \bm{\phi})$, $\rho_{ij}^{aa} = \rho(\|\bm{s}_i^a - \bm{s}_j^a\|; \bm{\phi})$, and $\rho_{ij}^{bb} = \rho(\|\bm{s}_i^b - \bm{s}_j^b\|; \bm{\phi})$, for $i, j = 1, \ldots, n/2$.

We consider the random vectors $\bm{U}_{ij} = \bm{Z}_i \big| \bm{Z}_j$, for $i\neq j$. This is the reason we refer to this method as bi-conditional likelihood (bi-CL). To derive concise expressions for the distribution of $\bm{U}_{ij}$, it is worth noting that $(\bm{Z}_i^\top,\bm{Z}_j^\top)^\top$ follows a quadri-variate Gaussian distribution, with zero mean, and covariance matrix
\[
 \begin{bmatrix}
     \bm{\Sigma}_{ii} & \bm{\Sigma}_{ij} \\
     \bm{\Sigma}_{ji} & \bm{\Sigma}_{jj}
  \end{bmatrix}  =  \begin{bmatrix}
\begin{array}{cc|cc}
\sigma^2 + \tau^2 & \sigma^2 \rho_{ii}^{ab} & \sigma^2 \rho_{ij}^{aa} & \sigma^2 \rho_{ij}^{ab} \\
& \sigma^2 + \tau^2 & \sigma^2 \rho_{ij}^{ba} & \sigma^2 \rho_{ij}^{bb} \\ \hline
& & \sigma^2 + \tau^2 & \sigma^2 \rho_{jj}^{ab} \\
& & & \sigma^2 + \tau^2 \\
\end{array}
\end{bmatrix}.
\]
We have excluded the values below the main diagonal, considering the matrix's symmetry. Consequently, $\bm{Z}_i \big| \bm{Z}_j=\bm{z}_j$ follows a bivariate Gaussian distribution with a mean of $\bm{\Sigma}_{ij} \bm{\Sigma}_{jj}^{-1} \bm{z}_j$ and a covariance matrix of $\bm{\Sigma}_{ii} - \bm{\Sigma}_{ij} \bm{\Sigma}_{jj}^{-1} \bm{\Sigma}_{ji}$, facilitating the derivation of its log-likelihood function $\ell_{ij}^{\text{(bi)}}(\tau^2,\sigma^2,\bm{\phi})$ (the explicit expression is defer to Appendix \ref{app:bicl} for a clearer exposition).  

The objective function for the bi-CL method is constructed following the conventional additive weighted methodology:
$$\text{\textcolor{black}{bi-cl}}(\tau^2,\sigma^2,\bm{\phi}) = \sum_{i=1}^{n/2} \sum_{j\neq i}^{n/2} \omega_{ij}  \ell_{ij}^{\text{(bi)}}(\tau^2,\sigma^2,\bm{\phi}),$$
\textcolor{black}{where $\omega_{ij}$, for $i,j\in\{1,\hdots,n/2\}$ and $i\neq j$, are weights.}
By interlacing bi-dimensional blocks, we anticipate achieving a potentially more robust integration of statistical information compared to the  PCL method. 

In addition, it is straightforward to see that $\ell_{ij}^{\text{(bi)}}$ can be expressed as the difference between the log-likelihood of the vector $(Z_i^a,Z_i^b,Z_j^a,Z_j^b)^\top$ and the log-likelihood of the pair $(Z_j^a,Z_j^b)^\top$. This implies that the objective function of bi-CL is an additive combination of two functions originating from the method developed by \cite{eidsvik2014estimation}.

The bi-CL method significantly relies on the design of blocks. In this study, we adopt a strategy that entails grouping spatial sites in proximity to each other. \textcolor{black}{Specifically, we consider the following steps:
\begin{itemize}
    \item we simulate $n/2$ points uniformly distributed in the region of interest;
    \item we then gradually form blocks by grouping the two nearest neighbor locations to a given simulated point.
\end{itemize}
}
The design may vary across experiments, depending on the fixed order of simulated points that determines the grouping sequence. Nevertheless, despite this variability, the architecture consistently approximates the desired proximity features.

\begin{figure}
    \centering
\includegraphics[scale=0.6]{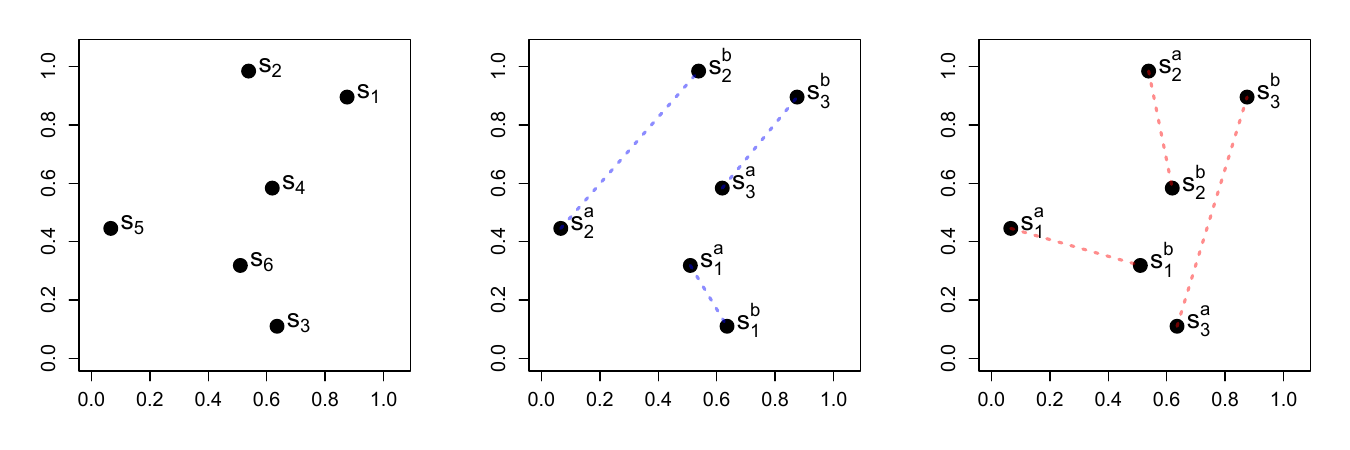}
    \caption{\textcolor{black}{From left to right: six spatial sites and two possible configurations with three bi-dimensional blocks. A dashed line connecting two points indicates that they form a bi-dimensional block.} }
    \label{fig:configurations}
\end{figure}

\textcolor{black}{In Figure~\ref{fig:configurations}, we illustrate this process for six sites, originally labeled as $\bm{s}_1,\ldots,\bm{s}_6$ (see the first panel). Following this scheme, we form three blocks of size two, labeled according to the notation introduced above: the first pair $(\bm{s}_1^a,\bm{s}_1^b)$, the second pair $(\bm{s}_2^a,\bm{s}_2^b)$, and the third pair $(\bm{s}_3^a,\bm{s}_3^b)$ (as shown in the second panel). We emphasize that this grouping depends on the random pairing scheme. Therefore, repeating the experiment yields alternative configurations of pairs (third panel). When combining pairs of blocks through conditioning, we consider, for example,
\[
\bm{Z}_1 \mid \bm{Z}_2 
= \big(Z(\bm{s}_1^a), Z(\bm{s}_1^b)\big) 
\mid \big(Z(\bm{s}_2^a), Z(\bm{s}_2^b)\big).
\]
In the first configuration (second panel in Figure~\ref{fig:configurations}), this corresponds to 
\[
\big(Z(\bm{s}_6), Z(\bm{s}_3)\big) 
\mid \big(Z(\bm{s}_5), Z(\bm{s}_2)\big),
\]
whereas in the second configuration (third panel in Figure~\ref{fig:configurations}), it becomes
\[
\big(Z(\bm{s}_5), Z(\bm{s}_6)\big) 
\mid \big(Z(\bm{s}_2), Z(\bm{s}_4)\big).
\]
Hence, the objective function depends on how the sites are paired. Since this pairing is not unique, we recommend generating multiple configurations and aggregating their corresponding objective functions additively. This results in a more stable objective function that captures a broader range of interactions among nearby observations.
}

For the weights, we will consider the following strategy, which is similar to the weights previously discussed for PCL:
$$\omega_{ij} = 
\begin{cases}
    1 & \text{if } \|\bm{s}_i^a-\bm{s}_j^a\| < d_s\\
    0 & \text{otherwise}
\end{cases},$$
\textcolor{black}{for $i,j\in\{1,\hdots,n/2\}$ and $i\neq j$}, and some positive constant $d_s$. Notice that we compare the first element of each block, and if their distance is below a given threshold, this pair enters the objective function. 

Note that the bi-CL method provides additional degrees of freedom, allowing for the creation of even more intricate weights. For instance, one could define weights based on factors such as maximum, minimum, or average element-wise distance between blocks. Nevertheless, we adhere to the aforementioned choice for its simplicity. Given the proposed strategy for grouping sites, we anticipate that any of these alternatives are expected to result in similar performances of the method.

In the GitHub repository \url{https://github.com/alfredoalegria/bi_CL}, we provide R scripts and C routines to perform bi-CL estimation, \textcolor{black}{including a fully documented illustrative example. This material focuses on illustrating the implementation of the block partitioning and the estimation procedure.}

\section{Simulation Studies}
\label{sec:simulations}

In this section, we assess the statistical and computational efficiency of diverse block likelihood estimation methods across different scenarios. This encompasses an examination of small- and large-sized blocking structures in comparison to the full likelihood. 

From now on, we will exclusively use the acronym BCL to refer to the block likelihood with large blocks, distinguishing it from PCL and bi-CL, each having its own designated acronyms.

\subsection{Simulation Setup}

We will focus on the following families of correlation functions:
\begin{itemize}
    \item Exponential model: $\displaystyle \rho(h; \phi) = \exp\left( -\frac{3\, h}{\phi} \right)$.
    \item Mat\'ern (1.5) model: $\displaystyle \rho(h; \phi) = \exp\left( - \frac{4.7619 \, h}{\phi} \right) \left( 1 +  \frac{4.7619 \, h}{\phi} \right)$.
 \item Cauchy model: $\displaystyle \rho(h; \phi)  = \left(1 + \left(\frac{4.3588 \, h}{\phi}\right)^2 \right)^{-1}$.
\end{itemize}
All models have been parameterized in terms of a range parameter $\phi$; that is, the correlation is smaller than 0.05 when $h > \phi$.  Thus, in all our experiments, the parameter vector is $\bm{\theta} = (\tau^2,\sigma^2,\phi)^\top$. 
The first two models are special cases of the Matérn family of correlation functions, with shape parameters of 0.5 and 1.5, respectively. This parameter governs the degree of smoothness of the correlation function at the origin, playing a crucial role in characterizing the level of differentiability (in a mean square sense) of the process (see, e.g., \citealp{stein2012interpolation}). Hence, the realizations from the Mat\'ern (1.5) model exhibit greater smoothness compared to those generated by the exponential model. On the other hand, the Cauchy model is characterized by polynomial decay, in contrast to Mat\'ern-type models, which exhibit exponential decay. This polynomial decay allows for a more persistent correlation between observations even when separated by large distances.

We consider the spatial domain $[0,1]^2$ and the following values for the range: $0.10$, $0.15$, and $0.20$. In all our experiments, we set $\sigma^2=1$ and $\tau^2=0.1$. 
  We will be working with a set of $n=500$ irregularly distributed sites. This moderate sample size allows for comparisons with the full likelihood. The generation of these sites adheres to the method outlined in \cite{bevilacqua2015comparing}: initially creating a regular grid with increments of $0.03$ over $[0,1]^2$, we then perturb each location by adding a uniform random value from $[-0.01,0.01]$ to each coordinate. Finally, we randomly select 500 points without replacement.

To maximize the objective functions, we employed a Nelder–Mead algorithm, allowing a maximum of $10^4$ iterations and a convergence tolerance of $10^{-16}$ between successive iterates. Given that we estimate three parameters, these specifications proved sufficient to attain accurate results for all scenarios; in fact, the number of iterations was only a few hundreds for all the experiments, and the method consistently converged to a solution without encountering any issues of degeneracy.

\subsection{Bi-CL vs. PCL}

We benchmark bi-CL against its closest competitor in the block likelihood range, the PCL method, both being approaches based on small-sized blocks. Both methods are implemented with 0/1 weights, where the threshold $d_s$  takes values of 0.05, 0.10 and 0.15. We added five configurations to implement the bi-CL method. 

We assess their effectiveness across the aforementioned scenarios and models, in comparison to the full likelihood (ML). The assessment involves computing relative root mean square errors for each parameter through Monte Carlo simulations, with 1{,}000 independent repetitions. We also assess the overall effectiveness using a global measure:
$$ \left( \frac{|\bm{G}_{\text{ML}}|^{1/2}}{|\bm{G}_{\text{CL}}|^{1/2}} \right)^{1/3},$$
where  $$\bm{G}_{\text{ML}} = \frac{1}{1{,}000}\sum_{i=1}^{1{,}000} \big(\hat{\bm{\theta}}_i^{\text{ML}} - {\bm{\theta}}\big) \big(\hat{\bm{\theta}}_i^{\text{ML}} - {\bm{\theta}}\big)^\top$$ \textcolor{black}{is the sample covariance matrix of the ML estimates}. The \textcolor{black}{sample covariance} matrices associated with the PCL and bi-CL methods, denoted in general here as $\bm{G}_{\text{CL}}$, are defined in a similar fashion. The metric involves the cubic root, as there are three parameters. 
\textcolor{black}{Since in the multi-parameter case variability is described by matrices rather than scalar variances, the determinant provides a natural scalar summary. This criterion is closely related to the notion of asymptotic relative efficiency, which is typically defined through ratios of determinants of information matrices (see, e.g., Equation~(23) in \citealp{bevilacqua2015comparing}).}

The results are displayed in Tables \ref{tab:rmse_exp}, \ref{tab:rmse_matern} and \ref{tab:rmse_cauchy}, for the exponential, Mat\'ern (1.5) and Cauchy models, respectively. In all scenarios, the bi-CL method consistently outperforms PCL. In the case of the exponential model, both methods demonstrate good performance. For example, when the range is 0.1 and the methods are implemented with $d_s=0.1$, PCL demonstrates a global performance of $91.03\%$, while bi-CL achieves $93.87\%$.  Observe that PCL exhibits poor performance when $d_s=0.05$ (i.e., when a drastically sparse weighting scheme is adopted), while bi-CL remains more stable. The differences between both methods become more pronounced for the Matérn (1.5) and Cauchy models. For instance, when the range is 0.1, the efficiency of bi-CL is $90.85\%$ in the Matérn (1.5) class, while the efficiency of PCL decreases to $83.98\%$. Similar findings are obtained in the other investigated scenarios.

In general, we observe a gradual degradation in the performance of these methods as the range increases, especially notable in the Matérn (1.5) and Cauchy models. In all scenarios, augmenting $d_s$ to 0.15 adversely affects the estimates of both methods. These results align with existing literature, particularly highlighting the detrimental consequences associated with an excess of terms in the objective function \citep{bevilacqua2015comparing}. Notably, when we consider the optimal weighting choice for each method, bi-CL consistently achieves a global efficiency of $85\%$ or higher in all studied scenarios, while this percentage drops to $76\%$ for PCL.

\begin{table}
\caption{Comparison of the relative root mean square error and relative global efficiency of the estimates with respect to the full likelihood, computed from 1{,}000 independent replicates for the exponential model.}
\label{tab:rmse_exp}
\centering
\begin{tabular}{ccccccccccc} \hline \hline 
\multicolumn{2}{c}{}     & &  \multicolumn{2}{c}{$d_s=0.05$} &  &  \multicolumn{2}{c}{$d_s=0.10$}  &  &  \multicolumn{2}{c}{$d_s=0.15$}  \\ 
 Range &  &   & {bi-CL} & {PCL} &  & {bi-CL} & {PCL} &  & {bi-CL} & {PCL} \\ 
\hline     

    & $\sigma^2$ && 0.8671 & 0.7910 && 0.9515 & 0.9202 && 0.9302 & 0.8568 \\
 0.10  & $\phi$  && 0.7353 & 0.2023 && 0.9051 & 0.8501 && 0.8625 & 0.7419 \\
    & $\tau^2$   && 0.8406 & 0.7436 && 0.9340 & 0.9034 && 0.8977 & 0.8183 \\
 & Global &&  0.8386 & 0.4522 && \bf{0.9387} & 0.9103 && 0.9107 & 0.8630 \\
      \hline

& $\sigma^2$ && 0.8827 & 0.7979 && 0.9593 & 0.9231 && 0.9501 & 0.8745 \\
0.15  & $\phi$  && 0.8006 & 0.4303 && 0.8949 & 0.8524 && 0.8496 & 0.7521 \\
& $\tau^2$   && 0.8484 & 0.7029 && 0.9244 & 0.8865 && 0.8863 & 0.7908 \\
   & Global &&    0.8553 & 0.6206 && \bf{0.9301} & 0.9025 && 0.8986 & 0.8534 \\
         \hline
         
    & $\sigma^2$ && 0.8809 & 0.8062 && 0.9507 & 0.9191 && 0.9423 & 0.8872 \\
 0.20  & $\phi$ && 0.8338 & 0.4718 && 0.8910 & 0.8601 && 0.8466 & 0.7652 \\
    & $\tau^2$   && 0.8469 & 0.6751 && 0.9196 & 0.8785 && 0.8763 & 0.7682 \\
  & Global && 0.8562 & 0.6248 && \bf{0.9227} & 0.8979 && 0.8885 & 0.8406 \\
\hline
\end{tabular}
\end{table}

\begin{table}
\caption{Comparison of the relative root mean square error and relative global efficiency of the estimates with respect to the full likelihood, computed from 1{,}000 independent replicates for the Mat\'ern (1.5) model.}
\label{tab:rmse_matern}
\centering
\begin{tabular}{ccccccccccc} \hline \hline 
\multicolumn{2}{c}{}     & &  \multicolumn{2}{c}{$d_s=0.05$} &  &  \multicolumn{2}{c}{$d_s=0.10$}  &  &  \multicolumn{2}{c}{$d_s=0.15$}  \\
 Range &  &   & {bi-CL} & {PCL} &  & {bi-CL} & {PCL} &  & {bi-CL} & {PCL} \\ 
\hline     

    & $\sigma^2$ && 0.8893 & 0.8329 && 0.9461 & 0.8888 && 0.9273 & 0.8348 \\
 0.10  & $\phi$ && 0.8439 & 0.7451 && 0.8851 & 0.7711 && 0.8403 & 0.6441 \\
    & $\tau^2$   && 0.8677 & 0.7737 && 0.8879 & 0.8031 && 0.8291 & 0.6919 \\
 & Global && 0.8624 & 0.8024 && \bf{0.9085} & 0.8398 && 0.8689 & 0.7782 \\
      \hline    

    & $\sigma^2$ && 0.8668 & 0.8064 && 0.9253 & 0.8866 && 0.9175 & 0.8697\\
 0.15  & $\phi$  && 0.8025 & 0.6781 && 0.8319 & 0.7428 && 0.7768 & 0.6153 \\
    & $\tau^2$   && 0.7716 & 0.6069 && 0.8080 & 0.6850 && 0.7280 & 0.5424 \\
 & Global &&    0.8165 & 0.7219 && \bf{0.8676} & 0.7840 && 0.8162 & 0.7126 \\
         \hline
   
    & $\sigma^2$ && 0.8601 & 0.8148 && 0.9073 & 0.8767 && 0.9023 & 0.8787 \\
 0.20  & $\phi$  && 0.7804 & 0.6432 && 0.8018 & 0.7350 && 0.7448 & 0.6148 \\
    & $\tau^2$   && 0.7385 & 0.5371 && 0.7897 & 0.6620 && 0.7020 & 0.5011 \\
& Global && 0.7956 & 0.6810 && \bf{0.8514} & 0.7685  && 0.7983 & 0.6906 \\
\hline
\end{tabular}
\end{table}

\begin{table}
\caption{Comparison of the relative root mean square error and relative global efficiency of the estimates with respect to the full likelihood, computed from 1{,}000 independent replicates for the Cauchy model.}
\label{tab:rmse_cauchy}
\centering
\begin{tabular}{ccccccccccc} \hline \hline 
\multicolumn{2}{c}{}     & &  \multicolumn{2}{c}{$d_s=0.05$} &  &  \multicolumn{2}{c}{$d_s=0.10$}  &  &  \multicolumn{2}{c}{$d_s=0.15$}  \\
 Range &  &   & {bi-CL} & {PCL} &  & {bi-CL} & {PCL} &  & {bi-CL} & {PCL} \\ 
\hline

    & $\sigma^2$ && 0.9234 & 0.9289 && 0.9485 & 0.9088 && 0.9212 & 0.8303 \\
 0.10  & $\phi$  && 0.8600 & 0.7229 && 0.8879 & 0.7411 && 0.8194 & 0.5579 \\
    & $\tau^2$   && 0.9203 & 0.9351 && 0.9274 & 0.8738 && 0.8967 & 0.7850 \\

   & Global &&   0.9017 &  0.7965 && \bf{0.9370} & 0.8743 && 0.8875 & 0.7543 \\
      \hline    

    & $\sigma^2$ && 0.9037 & 0.8496 && 0.9635 & 0.9220 && 0.9390 & 0.8437 \\
 0.15  & $\phi$  && 0.8716 & 0.8247 && 0.8382 & 0.6904 && 0.7686 & 0.5174 \\
    & $\tau^2$   && 0.9110 & 0.8717 && 0.8959 & 0.8142 && 0.8422 & 0.6666 \\
& Global &&  0.8807 & 0.8428 && \bf{0.8969} & 0.8255 && 0.8457 & 0.7216 \\
         \hline
         
    & $\sigma^2$ && 0.8604 & 0.7785 && 0.9335 & 0.8937 && 0.9202 & 0.8774 \\
 0.20  & $\phi$  && 0.8261 & 0.7660 && 0.7807 & 0.6427 && 0.7056 & 0.4861 \\
    & $\tau^2$   && 0.8401 & 0.7464 && 0.8156 & 0.7032 && 0.7411 & 0.5513 \\
& Global && 0.8368 & 0.7716 && \bf{0.8515} & 0.7645 && 0.7912 & 0.6713 \\
\hline
\end{tabular}
\end{table}

\textcolor{black}{In Tables~\ref{tab:rmse_exp}, \ref{tab:rmse_matern}, and \ref{tab:rmse_cauchy}, the global efficiencies initially increase as $d_s$ grows from 0.05 to 0.1, but then decline when $d_s$ increases further to 0.15. This pattern of initial improvement followed by decline aligns with the findings of \cite{bevilacqua2015comparing}[Figures 2 and 3], particularly regarding the estimation of the practical range.
Accordingly, the choice of $d_s$ is fundamental.
A practical and relatively robust approach is to select a value close to the practical range, which can be approximated by inspecting the empirical variogram of the data. }

\subsection{Small vs. Large Blocks}

We proceed to assess the performance of bi-CL and PCL against the BCL methods described in \cite{eidsvik2014estimation}. Despite all belonging to the same family, they represent opposing extremes within the block likelihood range.

\begin{figure}
    \centering
    \includegraphics[scale=0.115]{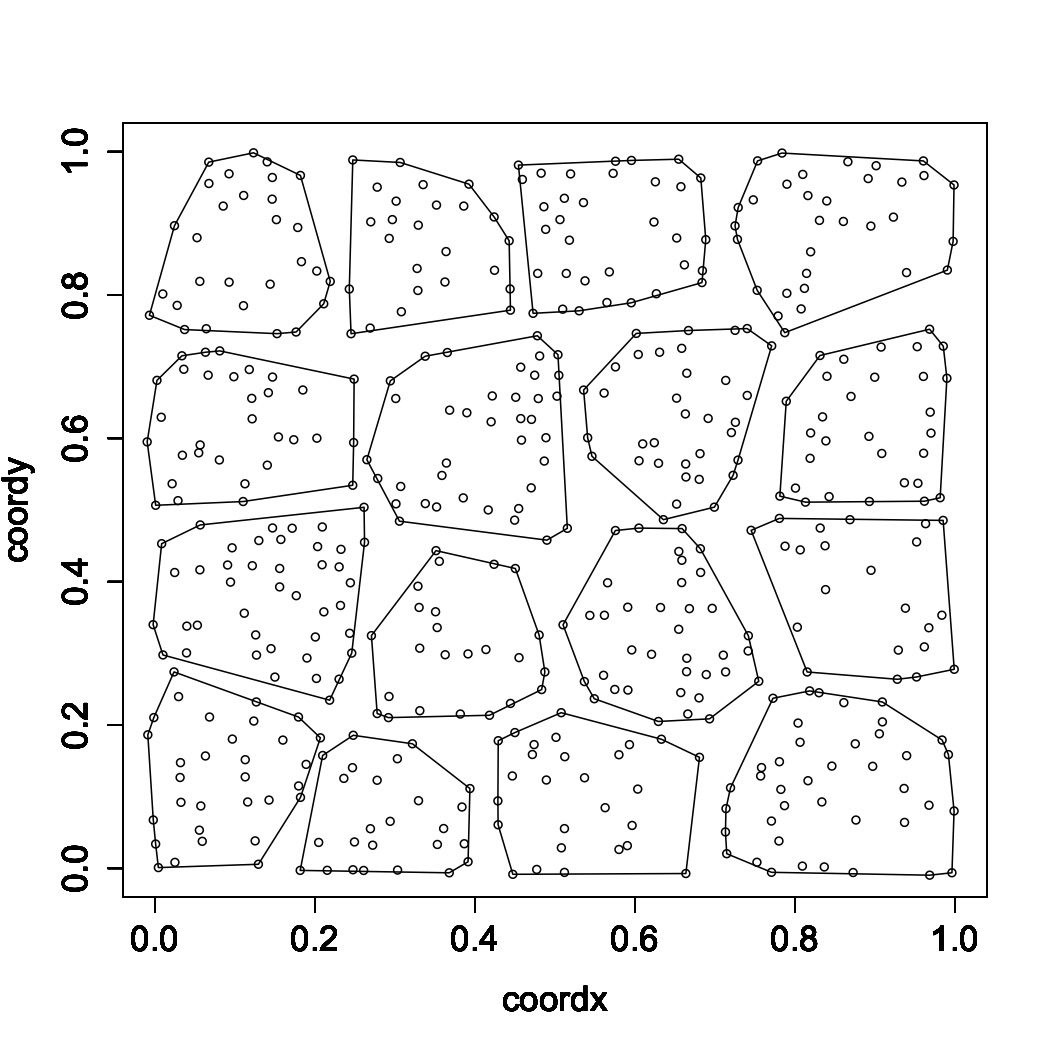} 
    \includegraphics[scale=0.115]{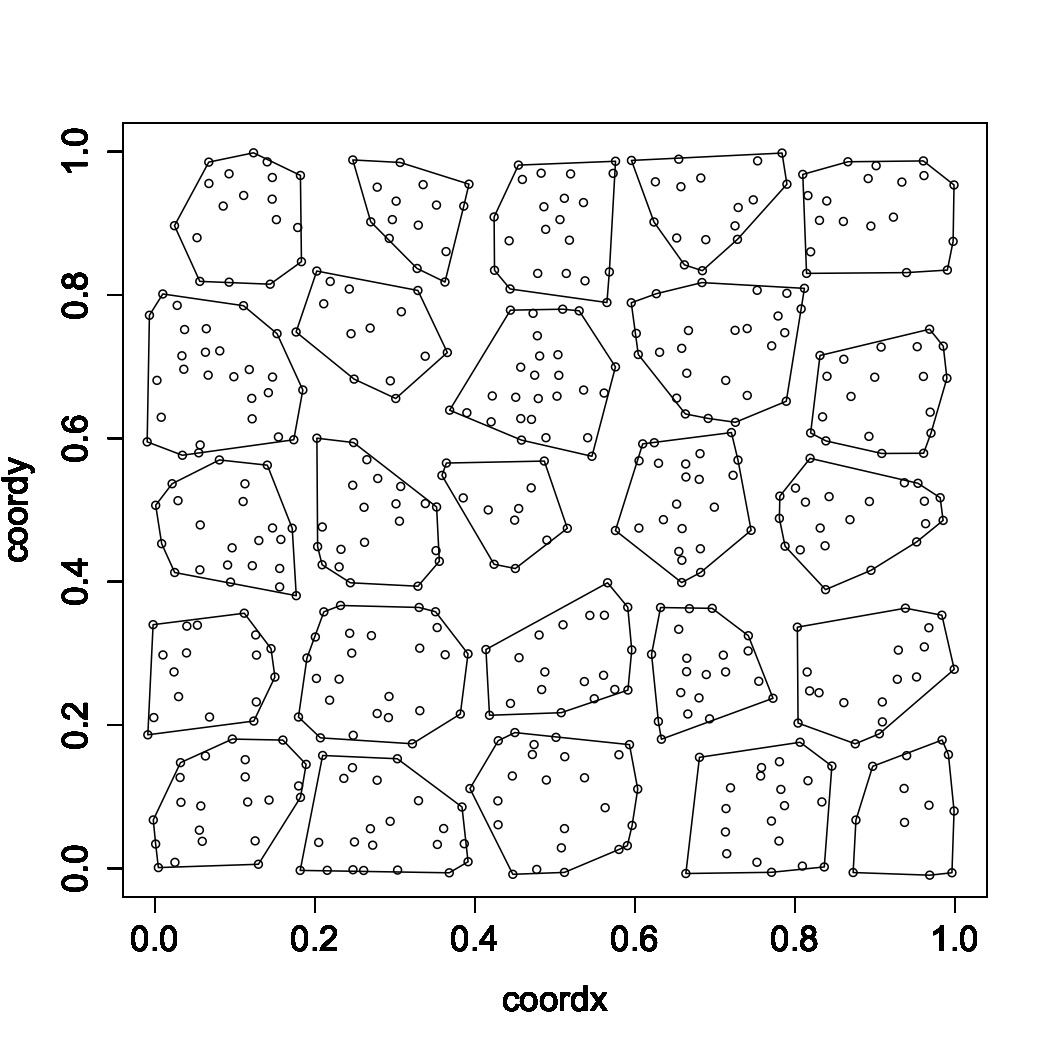} 
    \includegraphics[scale=0.115]{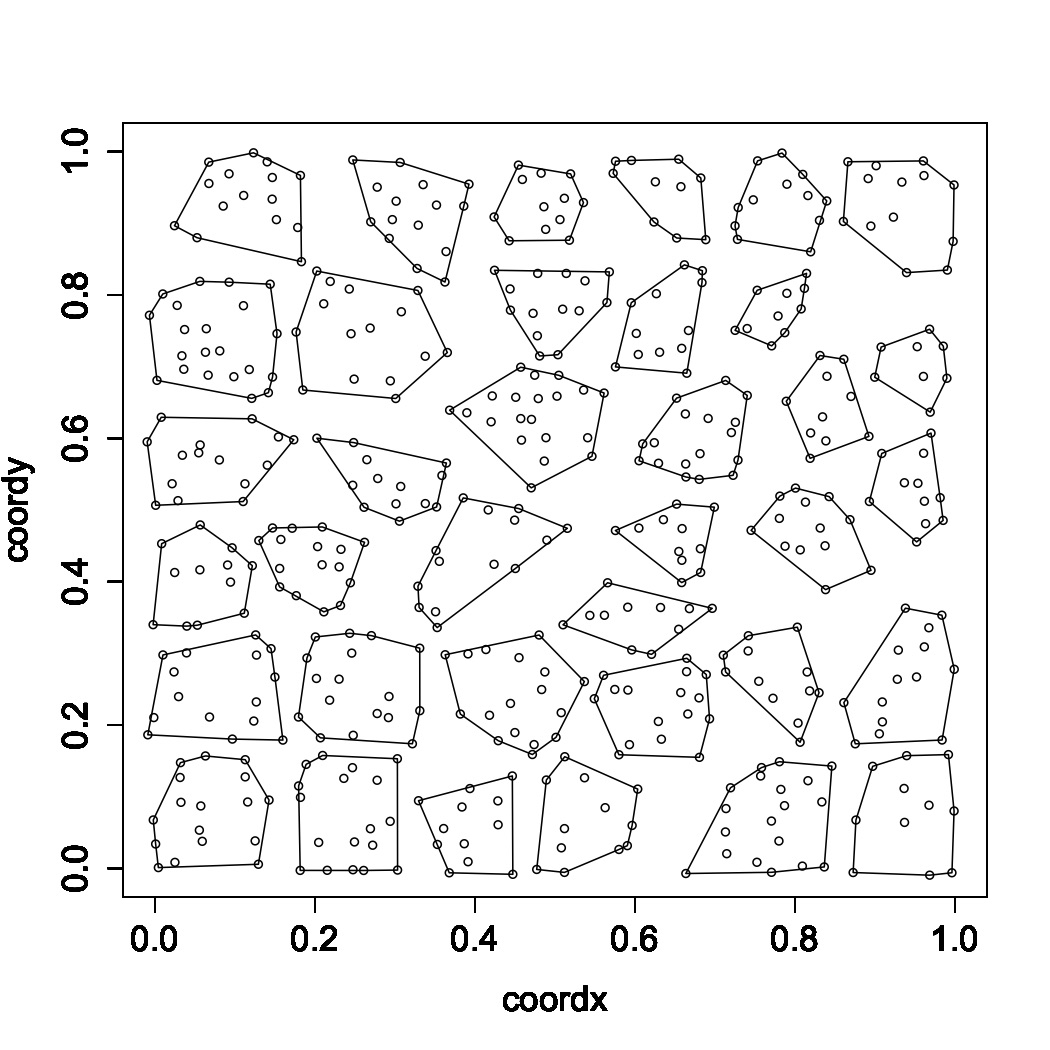}
    \caption{Simulated locations and convex hulls for block configurations with 16, 25, and 36 blocks (from left to right).}
    \label{fig:chull}
\end{figure}

In the same setting as the previous section, with a total of $n=500$ irregular sites within $[0,1]^2$, BCL is applied using $16$, $25$, and $36$ blocks. We use the notation BCL$_{16}$, BCL$_{25}$, and BCL$_{36}$ for these configurations, respectively. These blocks are constructed through clusters of sites (see Figure \ref{fig:chull}). It is worth noting that alternatives such as regular blocking, Voronoi tessellation, or even more intricate structures, are also possible \citep{eidsvik2014estimation}, but are not explored here. When forming pairs of blocks to construct the objective function, as detailed in Section \ref{sec:proposal}, we specifically consider blocks whose distances,  measured between their centers, are less than 0.3, 0.25, and 0.2 for BCL$_{16}$, BCL$_{25}$, and BCL$_{36}$, respectively. These criteria ensure that only neighboring blocks are taken into consideration.

\begin{figure}
    \centering
    \includegraphics[scale=0.25]{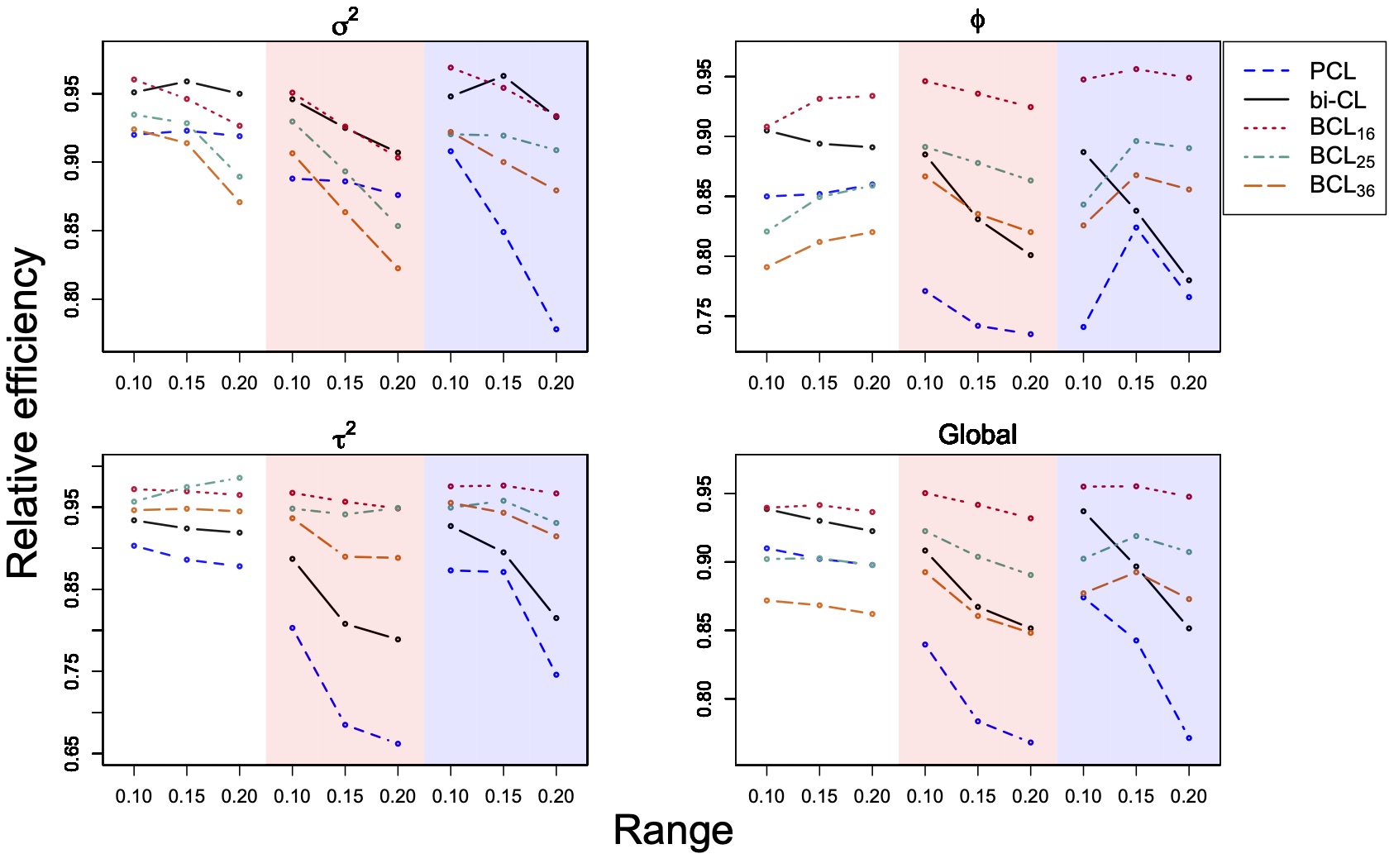}    
    \caption{Comparison of the relative root mean square error and relative global efficiency of the estimates with respect to the full likelihood, computed from 1{,}000 independent replicates. The three color zones in each panel correspond to the exponential, Mat\'ern (1.5) and Cauchy models (from left to right). BCL$_{16}$, BCL$_{25}$ and BCL$_{36}$ refer to block likelihood with 16, 25 and 36 blocks, respectively.}
    \label{fig:re_block}
\end{figure}

We assess the relative efficiency of BCL methods compared to ML, examining each parameter individually and employing a global measure, as described in the previous section. This enables us to contrast these efficiencies with those of bi-CL and PCL across the previously described scenarios. To provide a clearer perspective on each method's position in this comparison, the results are visually depicted in Figure \ref{fig:re_block}.

 The bi-CL method effectively bridges the gap between PCL and BCL methodologies. Specifically, for the exponential model, methods based on small-sized blocks prove notably effective, with bi-CL performing comparably to the best BCL method considered (BCL$_{16}$). In the Matérn (1.5) model, bi-CL is positioned between BCL$_{36}$ and BCL$_{25}$. Regarding the Cauchy model, the proposed method demonstrates strong performance when the range is small, comparable to BCL$_{16}$. However, as the range increases, the method experiences deterioration, mirroring the decline observed in PCL but consistently maintaining superiority over the latter and closely approximating the efficiency of BCL$_{25}$ and BCL$_{36}$. 
 
 These results reveal an interesting, and potentially counterintuitive, finding: bi-CL exhibits efficiencies comparable to, and in some cases exceeding, those of certain versions based on large-sized blocks.

\subsection{Timing Comparison}

We proceed to compare the methods in terms of their execution times. These experiments were carried out on a computer with 8GB of memory and a 1.1 GHz Intel Core i3 dual-core processor, ensuring accessibility and replicability for a diverse audience.

In this study, PCL is omitted as its times are found to be quite similar to those of bi-CL. Specifically for BCL, we quantify the time needed for Cholesky factorization, an essential step when dealing with large-sized blocks. We consider two scenarios where the sample is divided into either 8 or 16 blocks of similar size, resembling the cluster-based configurations in Figure \ref{fig:chull}. The bi-CL method is implemented with $d_s=0.1$.

In the case of 8 blocks, Cholesky factorization must be applied to matrices of size $n/4 \times n/4$ and computed at least 8 times, especially when the sites are irregularly spaced. This minimum count occurs when each block is paired with its closest counterpart. Similarly, when working with 16 blocks, Cholesky factorization is required for matrices of size $n/8 \times n/8$ and should be performed at least 16 times.

\begin{figure}
    \centering
    \includegraphics[scale=0.22]{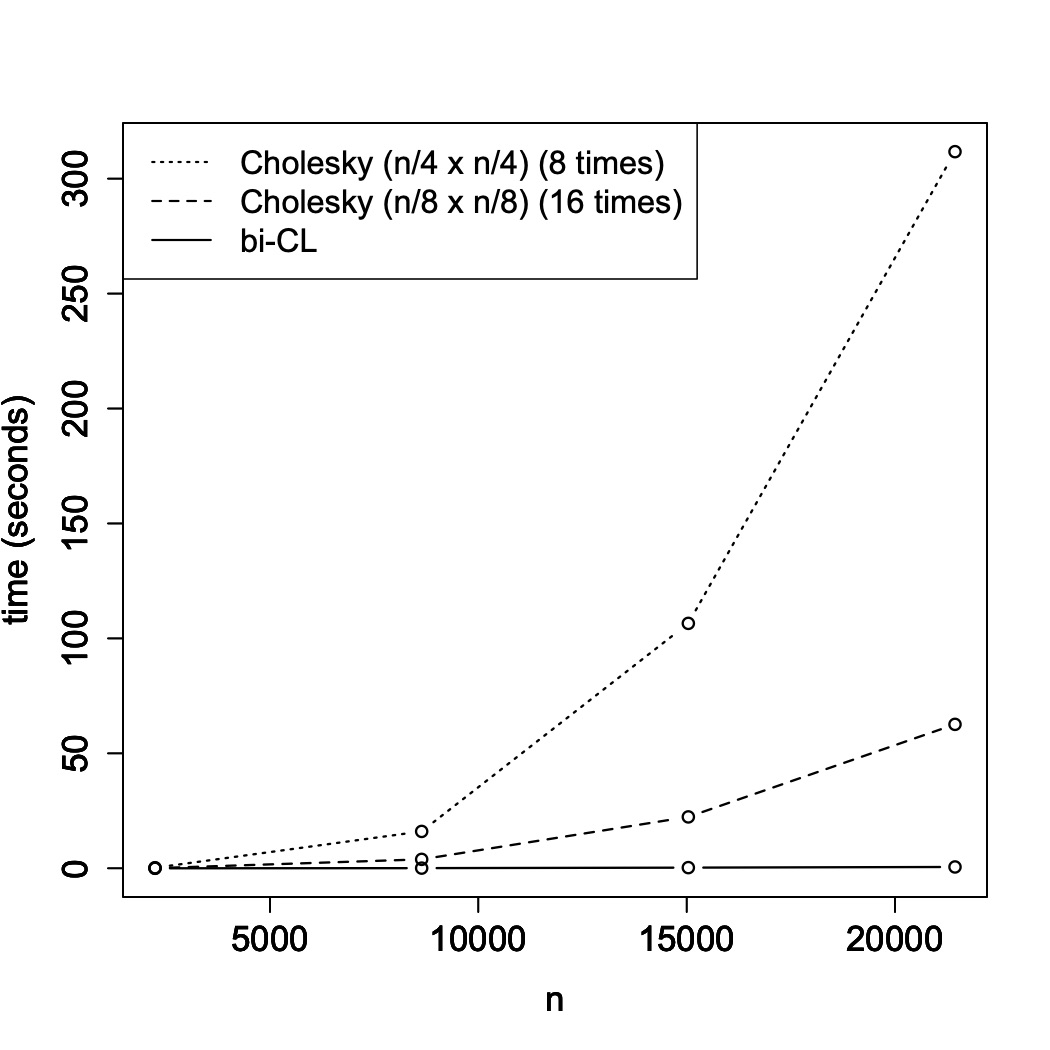}
    \caption{Execution time (in seconds) for evaluating the bi-CL method and the Cholesky factorizations required for BCL$_8$ and BCL$_{16}$ methods.}
    \label{fig:tiempos}
\end{figure}

The results are  displayed in Figure \ref{fig:tiempos}, considering $n$ ranging from 2,240 to 21,440. This experiment offers insight into the advantage of bi-CL in terms of execution time.  The evaluation of the objective function for the method with 16 blocks takes approximately 100 times longer than the evaluation for bi-CL, whereas for the method with 8 blocks, this figure increases to 520 times longer. As these times correspond to a single evaluation, this difference becomes more pronounced proportionally with the number of iterations when maximizing the objective function.

\textcolor{black}{
\subsection{Number of Configurations}
We conduct an additional simulation experiment to examine how the number of configurations, associated to the construction of the bi-dimensional blocks, influences the statistical performance of bi-CL. 
\\
The same simulation setup described at the beginning of the section is employed. Performance is assessed as a function of the number of configurations included in the objective function. Specifically, for each covariance model, and parameters $(\tau^2,\sigma^2,\phi)=(0.1,1,0.1)$, we generate 1{,}000 independent realizations of a Gaussian process and estimate the parameters through bi-CL using 2, 4, 6, and 8 configurations. \\
Figure~\ref{fig:number_config} presents the corresponding boxplots of the estimates for each scenario. Across all cases, a slight reduction in estimation variance is observed as the number of configurations increases, with a more pronounced effect on the practical range and the nugget effect, particularly for the Mat\'ern $(1.5)$ and Cauchy covariance models. \\
As a guideline for practitioners, we recommend using between 5 and 10 configurations to achieve a stable performance in this experiment with $N=500$ sites. The number of configurations should increase gradually with the sample size.
\begin{figure}
    \centering
    \includegraphics[scale=0.55]{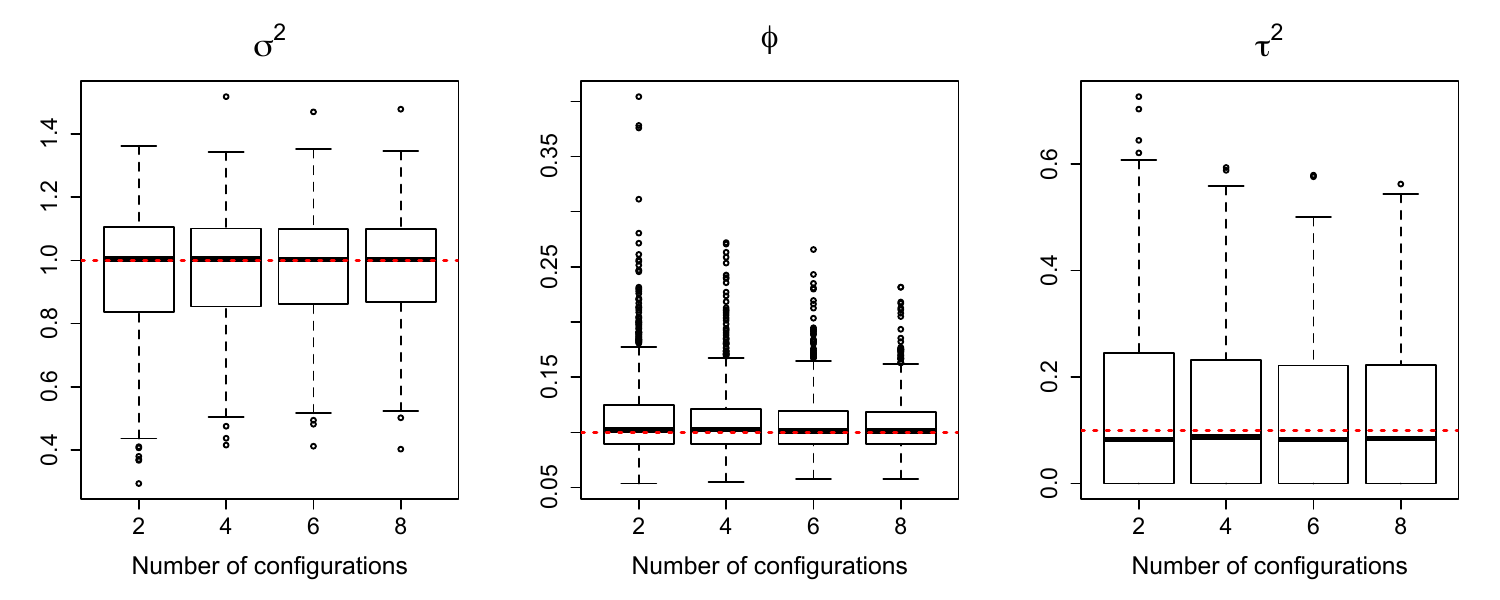}
     \includegraphics[scale=0.55]{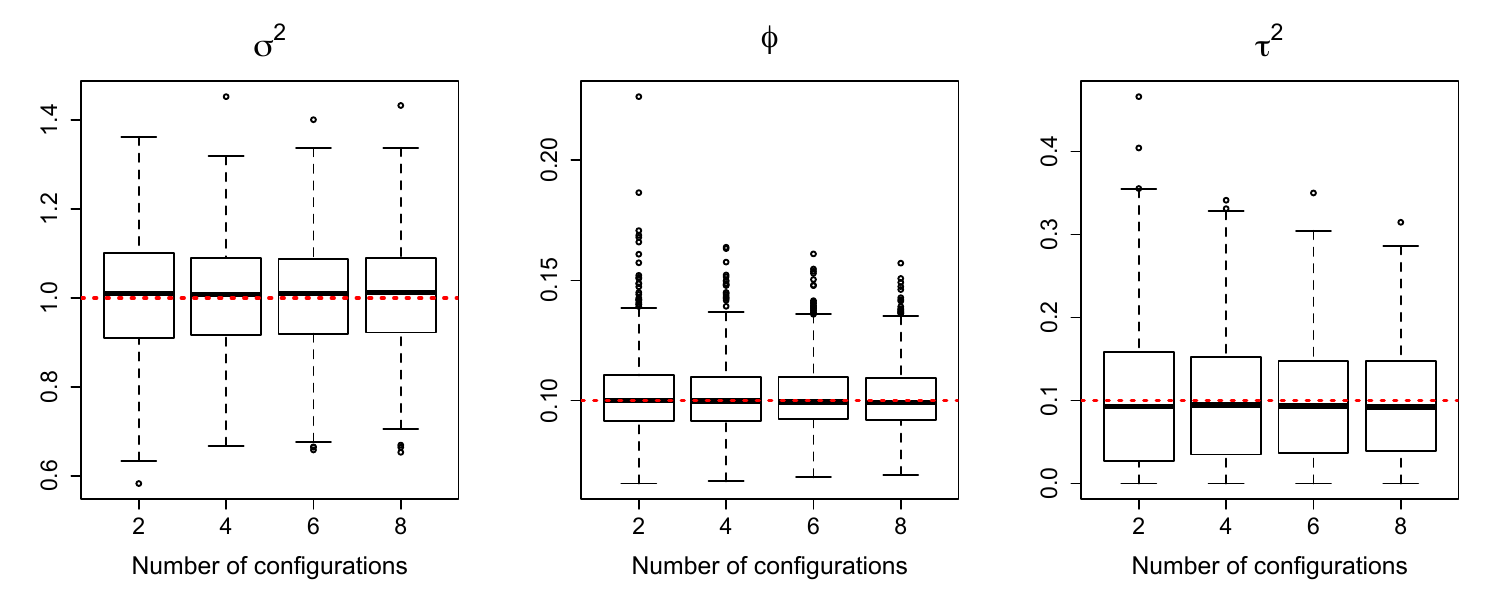}
      \includegraphics[scale=0.55]{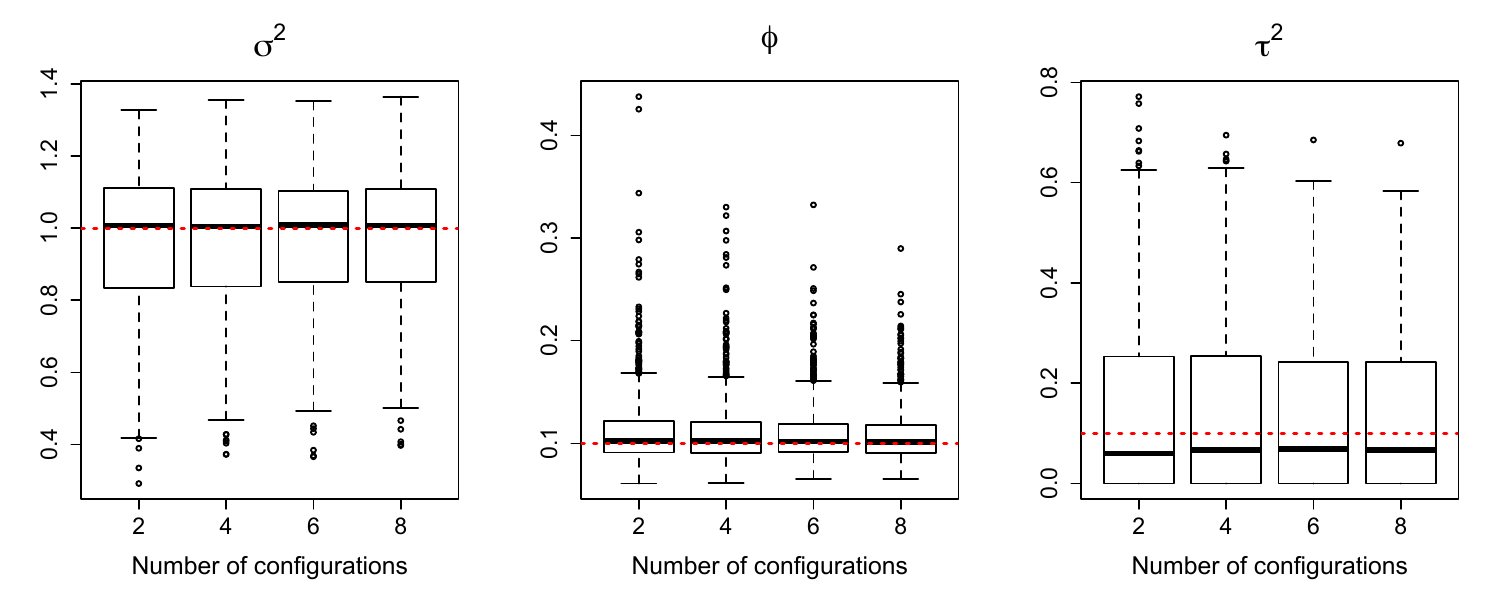}
    \caption{\textcolor{black}{Boxplots of bi-CL estimates as a function of the number of configurations included in the objective function, based on 1{,}000 independent realizations of a Gaussian process. From top to bottom: Exponential, Mat\'ern ($1.5$), and Cauchy covariance models. The dotted lines represent the true parameter values. }}
    \label{fig:number_config}
\end{figure}
}

\textcolor{black}{
\subsection{Asymmetric Weights}
This section extends the numerical comparison by incorporating the PCL approach with asymmetric weights, as described in Equation~(\ref{peso_asim}). We simulate 1{,}000 independent realizations of a Gaussian process with an Exponential covariance function and parameters $(\tau^2,\sigma^2,\phi)=(0.1,1,0.1)$, under the spatial design described at the beginning of this section. Parameter estimates are obtained using bi-CL and PCL with distance-based weights ($d_S = 0.05$). In addition, we consider PCL with asymmetric weights for $k=1,2,3$. 
\smallskip \\ 
Figure~\ref{fig:asymmetric} shows the corresponding boxplots of the estimates and suggests that asymmetric weighting can improve the performance of PCL in this setting. In particular, for $k = 3$, the results are comparable to those of the standard PCL with $d_S = 0.05$, but with reduced variability, as reflected by shorter whiskers and fewer outliers.
\begin{figure}
    \centering
    \includegraphics[scale=0.62]{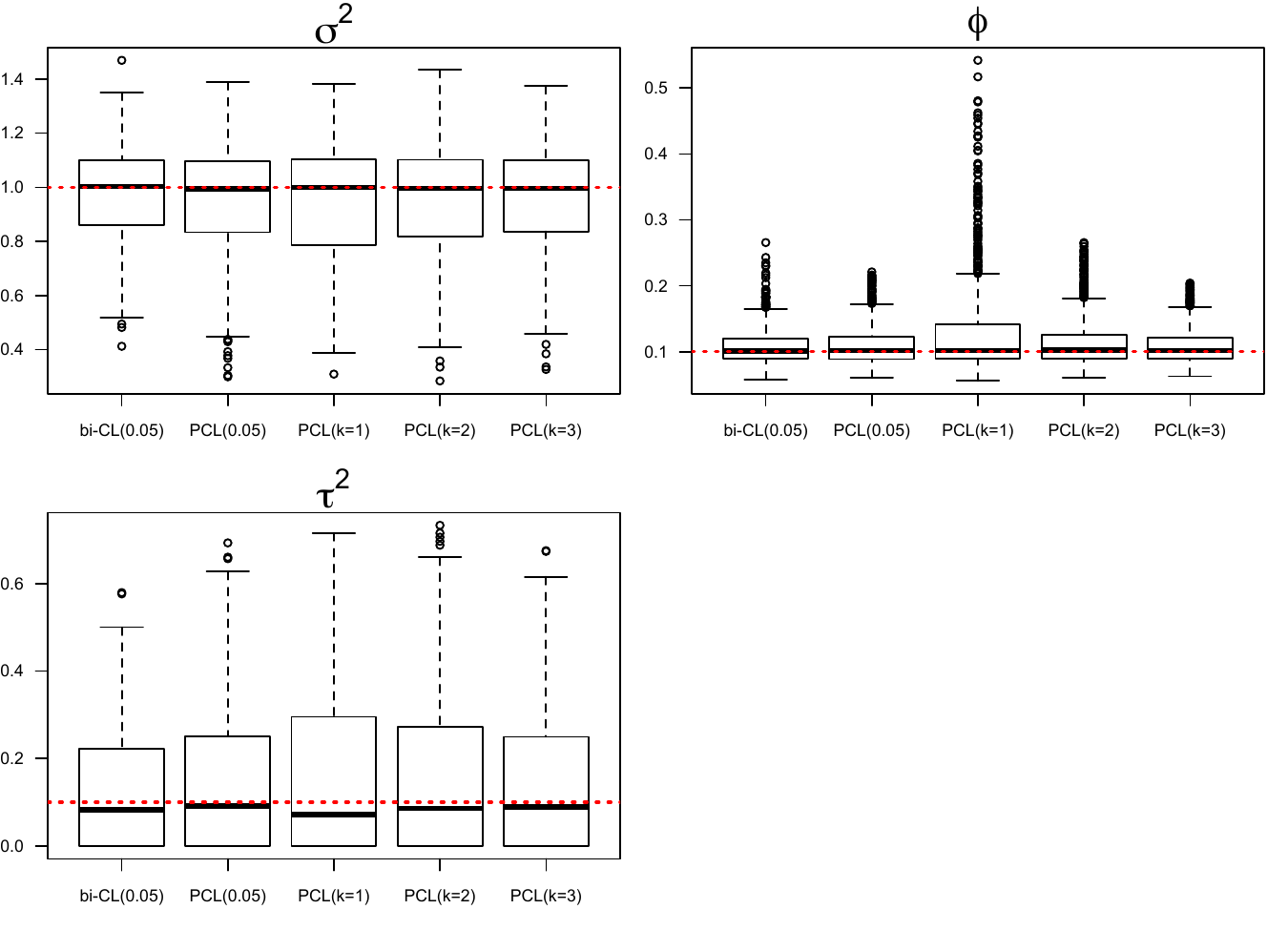}
    \caption{\textcolor{black}{Boxplots of estimates based on 1{,}000 independent realizations of a Gaussian process with Exponential covariance function. Results are shown for bi-CL and PCL with distance-based weights ($d_S = 0.05$), and for PCL with asymmetric weights for varying $k$. The dotted lines represent the true parameter values.} }
    \label{fig:asymmetric}
\end{figure}
\smallskip 
\\
These results reinforce the conclusions in \cite{caamano2024nearest}, where asymmetric weights emerge as an interesting alternative to conventional distance-based weighting schemes. The bi-CL method also has the potential to incorporate asymmetric weights. Since bi-CL relies on two-dimensional blocks, one could define neighboring blocks, for example, based on centroids.
 }

\section{Real Data Example}
\label{sec:data}

We examine the COBE Sea Surface Temperature data from NOAA PSL in Boulder, Colorado, accessible at \url{https://psl.noaa.gov} (see \citealp{https://doi.org/10.1002/joc.1169}). This dataset, with a spatial resolution of $1^\circ \times 1^\circ$, plays a fundamental role in tracking climate change trends, gaining insights into ocean circulation patterns, forecasting extreme weather incidents, and influencing policies for marine resource management, as well as reducing the vulnerability of communities to the effects of climate changes.

\begin{figure}
    \centering
    \includegraphics[scale=0.23]{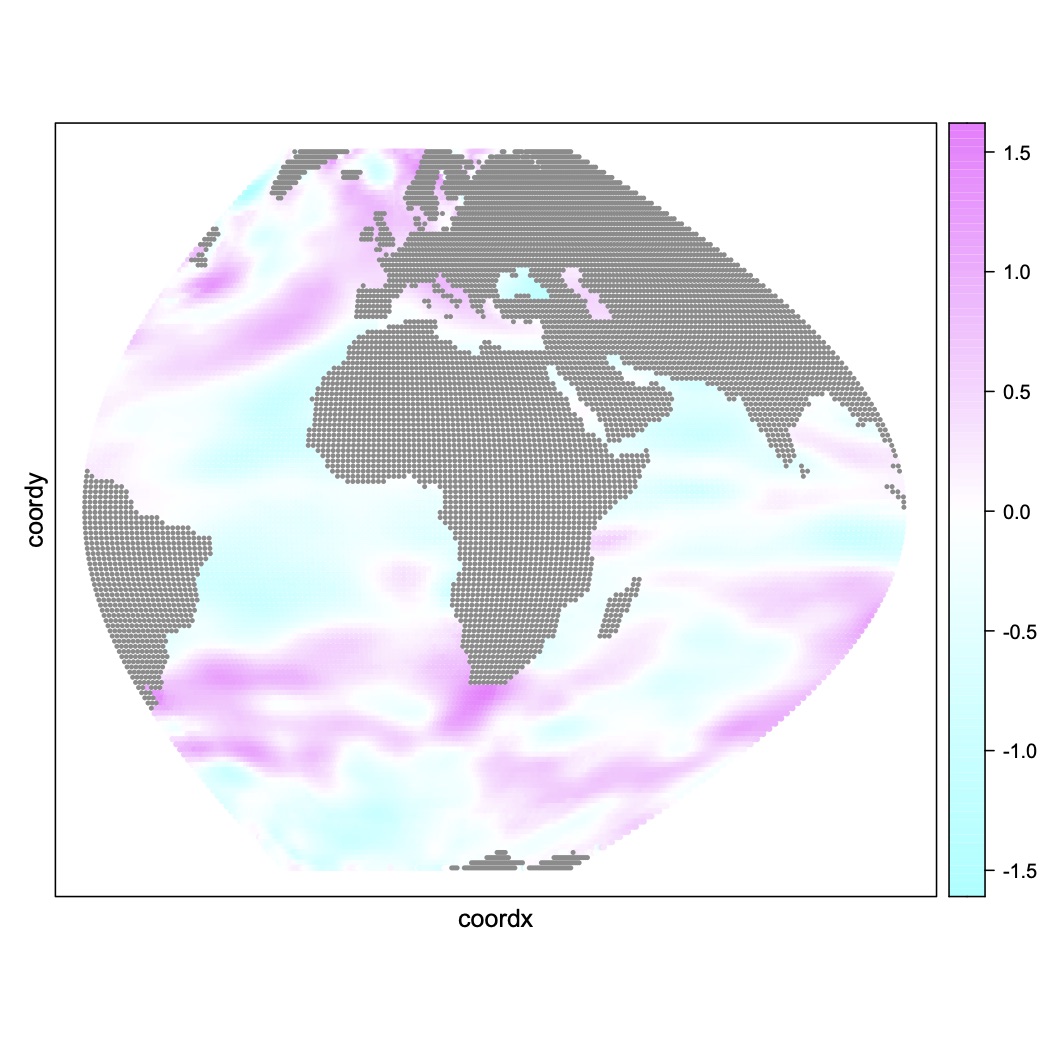}
    \caption{Sea surface temperature anomalies (in degrees Celsius) for March 2012.}
    \label{fig:sst}
\end{figure}

In our analysis, we focus on the mean anomalies for March 2012 within the study area covering the Atlantic and Indian Oceans. This area corresponds to latitudes between 70$^\circ$S and 70$^\circ$N and longitudes between 60$^\circ$W and 100$^\circ$E. This enhances the data characterization using an isotropic Gaussian process. Additionally, as a small remnant mean persists, we center the dataset by subtracting the sample mean, which is $0.00968$. The resulting dataset comprises 14,884 observations, with locations projected onto the plane using sinusoidal projection (refer to Figure \ref{fig:sst}).

Given the apparent smoothness in the variable of interest, we employ the Matérn (1.5) covariance function, and the model is augmented with a nugget effect. We estimate the parameters $(\tau^2, \sigma^2, \phi)$ using both the bi-CL and PCL methods, incorporating a spatial distance threshold for the weights set at $d_s = 500$ km. With a minimum distance of about 39 km between different points in the zone under study, this weighting scheme encompasses a reasonable number of pairs. As in the simulation experiments, the bi-CL method is implemented by adding 5 configurations. Additionally, we implement a large-sized BCL approach with 100 blocks. To ensure computational efficiency, not only for point estimation but also for subsequent standard error calculations, each block is exclusively paired with its nearest neighbor block.

\begin{figure}
    \centering
    \includegraphics[scale=0.25]{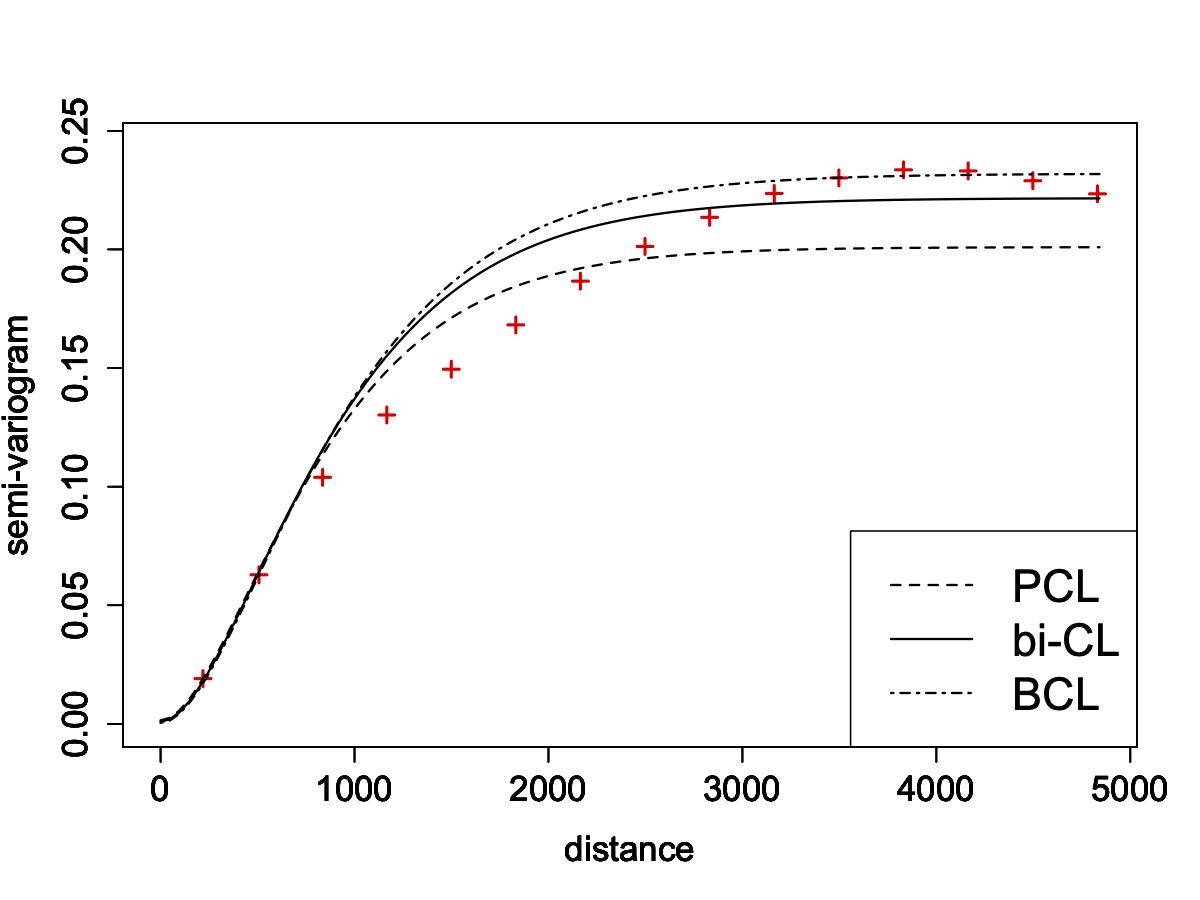}
    \caption{Sample (+) and fitted semi-variograms for sea surface temperature anomalies.}
    \label{fig:variogram}
\end{figure}

The results are presented in Table \ref{data_estimates}. Standard errors, enclosed in parentheses within Table \ref{data_estimates}, are computed using parametric bootstrap, generating 100 replicates based on the estimated models, and subsequently calculating sample standard deviations across these replicates. As discussed in \cite{bai2012joint}, although this strategy involves more computation, it is less susceptible to bias compared to alternatives based on subsampling techniques. Note that all methods identify an almost negligible nugget effect. While the estimates from all methods are comparable, PCL seems to slightly underestimate the variance and range parameters. Furthermore, PCL exhibits a larger standard error in the estimation of the range compared to its competitors. In contrast, the standard error of the variance estimate is smaller. Figure \ref{fig:variogram} depicts the sample and fitted semi-variogram for each method, demonstrating a reasonable alignment between theoretical and observed quantities.

We also evaluate the methods concerning predictive performance. Table \ref{data_estimates} illustrates the root mean squared error (RMSE) from a leave-one-out prediction study using simple kriging. The methods display similar results, with BCL performing the best, followed by bi-CL, and PCL showing slightly less favorable results.  Specifically, the predictive performance gap between PCL and BCL is $7.9\%$, and bi-CL reduces this gap by approximately $3\%$. These findings align with the results obtained from experiments using synthetic data, emphasizing the significant role of bi-CL, which positions itself with a level of accuracy between PCL and BCL methods with large blocks.

\begin{table}[]
    \caption{Parameter and standard error estimates for the Mat\'ern (1.5) model with a nugget effect fitted to the sea surface temperature anomalies dataset. The last column also reports the root mean square errors associated to the leave-one-out cross validation study.}
    \label{data_estimates}
    \centering
    \begin{tabular}{ccccc} \hline \hline
 & $\tau^2$  & $\sigma^2$  & $\phi$  & RMSE  \\ \hline
bi-CL & 0.000916 (0.000075)   & 0.2207 (0.0308) & 2,283 (143.64)  & 0.02978 \\
BCL   & 0.000438 (0.000012)   & 0.2314 (0.0274) & 2,381 (101.68)  & 0.02831 \\
PCL   & 0.001479 (0.000191)   & 0.1994 (0.0266) & 2,113 (164.19)  & 0.03077 \\
\hline
    \end{tabular}
\end{table}

\section{Discussion}
\label{sec:conclusions}

We have presented a comprehensive numerical comparison of block likelihood methods. We have identified key aspects:
\begin{itemize}
\item Our studies demonstrate the superior performance of bi-CL compared to the conventional PCL method, consistently revealing advantages across diverse scenarios. Significant improvements are observed, particularly for the Matérn (1.5) and Cauchy correlation structures. Importantly, these enhancements come without adding computational complexity. 
    \item  Regarding the exponential covariance model, there are no significant improvements in the  statistical efficiency  when employing large blocks. The bi-CL method proves to be highly competitive in this scenario, irrespective of the spatial range.
\item For the Matérn (1.5) and Cauchy covariance models, the bi-CL method exhibits favorable results, particularly when the range is small. It maintains a global statistical efficiency similar to large-sized blocks counterparts while achieving a better balance with respect to computational burden.

\item With an increase in the range, methods based on small-sized blocks experience a decline in statistical performance for the Matérn (1.5) and Cauchy covariance models. However, the bi-CL method continues to produce results that fall between the analyzed variants with large blocks,  unlike the pairwise likelihood method which consistently positions itself at the bottom.
\end{itemize}
These findings challenge the intuitive notion that universally increasing block size leads to improved performance.  Consequently, methods based on small blocks, especially bi-CL, emerge as a viable alternative. 

In this way, our study complements the works of \cite{bevilacqua2015comparing} and \cite{eidsvik2014estimation} by offering additional insights into the effectiveness of small-block strategies and examining a broader range of scenarios.

\textcolor{black}{In our analysis, we focus on scenarios with a zero mean for the sake of simplicity, primarily motivated by the fact that the data to which we apply our results consist of temperature anomalies. It is worth noting that, when the mean is non-constant or misspecified, the efficiency of covariance parameter estimates may be affected. Nevertheless, the framework can be readily extended to more general mean structures. In particular, one may consider a specification of the form $E\big(Z(\bm{s})\big) = \bm{x}(\bm{s})^\top \bm{\beta}$, where $\bm{x}(\bm{s})$ is a vector of spatially referenced explanatory variables and $\bm{\beta}$ is a vector of parameters, as in \cite{eidsvik2014estimation}. The methods studied here can then be applied in a similar manner.}

\textcolor{black}{Although the real application in Section \ref{sec:data}  considers a sample size of about 15{,}000, CL methods based on pairs have already been  used in settings involving hundreds of thousands of points (see, e.g., \citealp{bevilacqua2012estimating} for a spatio-temporal example), and, as bi-CL inherits the same computational order, it exhibits a comparable scalability profile.}

\textcolor{black}{Regarding comparisons with alternative approaches, the Vecchia approximation is an emerging gold standard within likelihood approximations in geostatistics. Under an optimal implementation (with carefully chosen neighbor sets and ordering), it has been shown to achieve performance comparable to the exact likelihood for Gaussian processes \citep{10.1214/19-STS755}. Consequently, our comparisons with the full likelihood can also be interpreted as an indirect benchmark against Vecchia-type methods under ideal conditions.}

\textcolor{black}{Importantly, the development of computationally efficient estimation methods in geostatistics is not a settled problem with a single optimal solution, so it is important to  consider the pros and cons of different approaches. In this context, BCL with small block sizes offers flexibility, particularly in non-Gaussian settings. For instance, in skew-Gaussian processes, the likelihood can be expressed as a combination of $2^n$ Gaussian likelihoods \citep{zhang2010spatial}, with $n$ denoting the sample size, while in wrapped-Gaussian processes it involves $n$ infinite series of Gaussian densities \citep{jona2012spatial}; after truncation to $k$ terms, this leads to expressions requiring the evaluation of $k^n$ Gaussian densities. In these cases, BCL becomes necessary, as working with small blocks remains computationally feasible, whereas methods relying on higher-order joint densities quickly become intractable. In this context, approaches such as Vecchia may be challenging to apply. In addition, Vecchia-type methods have been less  explored in high-dimensional spatial domains, whereas this does not pose a limitation for composite likelihood methods.}






\bmhead{Acknowledgements}

The author acknowledges the funding of the National
Agency for Research and Development of Chile, through Grant ANID Fondecyt 1251154.



\section*{Declarations}

\subsection*{Competing interests}
The author declares no competing interests.

\begin{appendices}
\section{Objective Function of Bi-CL}
\label{app:bicl}

\textcolor{black}{Recall that
\begin{equation}
\label{biv}
    \bm{U}_{ij} = \bm{Z}_i \mid \bm{Z}_j  \sim \mathcal{N}_2\big(  \bm{\mu} , \bm{\Xi} \big),
\end{equation}
where $\bm{\mu} = \bm{\Sigma}_{ij}  \bm{\Sigma}_{jj}^{-1} \bm{Z}_j$ and $\bm{\Xi} = \bm{\Sigma}_{ii} -\bm{\Sigma}_{ij}  \bm{\Sigma}_{jj}^{-1} \bm{\Sigma}_{ji}$.
Since $$\bm{\Psi} :=\bm{\Sigma}_{ij} \bm{\Sigma}_{jj}^{-1} = 
\begin{bmatrix}
    \sigma^2 \rho_{ij}^{aa} & \sigma^2 \rho_{ij}^{ab}\\
\sigma^2 \rho_{ij}^{ba} & \sigma^2 \rho_{ij}^{bb}
\end{bmatrix} 
\begin{bmatrix}
    \sigma^2 + \tau^2 & \sigma^2 \rho_{jj}^{ab}\\
\sigma^2 \rho_{jj}^{ab} &  \sigma^2 + \tau^2
\end{bmatrix}^{-1}
$$ 
appears in both the mean vector and the covariance matrix in (\ref{biv}), it is convenient to compute its entries explicitly. Using straightforward calculations for $2\times 2$ matrices, the elements of $\bm{\Psi}$ are given by
\begin{itemize}
\item  $\psi_{11} = \displaystyle \sigma^2 \left[ \frac{(\sigma^2+\tau^2)\rho_{ij}^{aa} - \sigma^2  \rho_{ij}^{ab} \rho_{jj}^{ab}  }{(\sigma^2+\tau^2)^2 - (\sigma^2 \rho_{jj}^{ab})^2}\right]$ \medskip  
\item $\psi_{21}  = \displaystyle \sigma^2 \left[\frac{(\sigma^2+\tau^2) \rho_{ij}^{ba} - \sigma^2 \rho_{ij}^{bb} \rho_{jj}^{ab} }{(\sigma^2+\tau^2)^2 - (\sigma^2 \rho_{jj}^{ab})^2} \right]$  \medskip  
    \item $\psi_{12} = \displaystyle \sigma^2 \left[ \frac{(\sigma^2+\tau^2) \rho_{ij}^{ab} - \sigma^2 \rho_{ij}^{aa} \rho_{jj}^{ab}}{(\sigma^2+\tau^2)^2 - (\sigma^2 \rho_{jj}^{ab})^2} \right]$ \medskip 
\item $\psi_{22} = \displaystyle \sigma^2 \left[ \frac{(\sigma^2+\tau^2) \rho_{ij}^{bb} - \sigma^2\rho_{ij}^{ba} \rho_{jj}^{ab}}{(\sigma^2+\tau^2)^2 - (\sigma^2 \rho_{jj}^{ab})^2} \right]$ \medskip
\end{itemize}
 \noindent These expressions allow us to compute the mean vector $\bm{\mu} = \bm{\Psi} \bm{Z}_j$ and the covariance matrix $\bm{\Xi} = \bm{\Sigma}_{ii} - \bm{\Psi}\bm{\Sigma}_{ji}$ in a more modular manner, again relying on standard $2\times 2$ matrix calculations. Hence, 
\begin{eqnarray*}
\ell_{ij}^{\text{(bi)}}(\tau^2,\sigma^2,\bm{\phi})  & = & -\frac{1}{2} \bigg(  \log |\bm{\Xi} | + (\bm{Z}_i-\bm{\mu})^\top \bm{\Xi}^{-1} (\bm{Z}_i-\bm{\mu})  \bigg)\\ 
& = & -\frac{1}{2}  \bigg(  \log(\xi_{11}\xi_{22} - \xi_{12}^2)   + \frac{\xi_{22} (Z_i^a-\mu_1)^2\ + \xi_{11} (Z_i^b-\mu_2)^2 - 2\xi_{12} (Z_i^a-\mu_1)  (Z_i^b-\mu_2)}{\xi_{11}\xi_{22} - \xi_{12}^2}\bigg),
\end{eqnarray*}
where 
\begin{itemize}
\item $\displaystyle \xi_{11} = \tau^2 + \sigma^2(1- \psi_{11}\rho_{ij}^{aa} - \psi_{12}\rho_{ij}^{ab})$
\item 
$\xi_{12} = \displaystyle    \sigma^2(\rho_{ii}^{ab} - \psi_{11}\rho_{ij}^{ba} - \psi_{12}\rho_{ij}^{bb})$
    \item  $\xi_{22} = \displaystyle   \tau^2  + \sigma^2(1-\psi_{21}\rho_{ij}^{ba} - \psi_{22}\rho_{ij}^{bb})$
\end{itemize} and
\begin{itemize}
   \item  $\displaystyle  \mu_1 = \psi_{11}Z_j^a + \psi_{12}Z_j^b $
\item $ \displaystyle  \mu_2 = \psi_{21} Z_j^a + \psi_{22} Z_j^b$
\end{itemize}
are precisely the elements of $\bm{\Xi}$ and $\bm{\mu}$, respectively. 
 }

\end{appendices}

\bibliography{mybib}

\end{document}